\documentclass[aps,prl,twocolumn,groupaddress,superscriptaddress]{revtex4-2}

\usepackage{graphicx}
\usepackage{multirow}
\usepackage{longtable}
\usepackage{stackengine}
\usepackage{amssymb,amsmath}
\newcommand\xrowht[2][0]{\addstackgap[.5\dimexpr#2\relax]{\vphantom{#1}}}

\usepackage{color, colortbl}

\definecolor{Gray}{rgb}{0.92, 0.92, 0.92}

\usepackage{float}
\usepackage{hhline}

\usepackage[table]{xcolor}

\begin{document}


\title{Diversity of Radial Spin Textures in Chiral Materials}


\author{Daniel Gos\'albez-Mart\'inez}
\email[]{daniel.gosalbez@ua.es}
\affiliation{Departamento de Física Aplicada, Universidad de Alicante, 03690 Alicante, Spain}
\affiliation{Institute of Physics, Ecole Polytechnique F\'ed\'erale de Lausanne (EPFL), CH-1015 Lausanne, Switzerland}
\affiliation{National Centre for Computational Design and Discovery of Novel Materials MARVEL, Ecole Polytechnique F\'{e}d\'{e}rale de Lausanne (EPFL), CH-1015 Lausanne, Switzerland}

\author{Alberto Crepaldi}
\affiliation{Dipartimento di Fisica, Politecnico di Milano, Piazza Leonardo da Vinci 32, Milan 20133, Italy}

\author{Oleg V. Yazyev}
\email[]{oleg.yazyev@epfl.ch}
\affiliation{Institute of Physics, Ecole Polytechnique F\'ed\'erale de Lausanne (EPFL), CH-1015 Lausanne, Switzerland}
\affiliation{National Centre for Computational Design and Discovery of Novel Materials MARVEL, Ecole Polytechnique F\'{e}d\'{e}rale de Lausanne (EPFL), CH-1015 Lausanne, Switzerland}


\date{\today}

\begin{abstract}
We introduce a classification of the radial spin textures in momentum space that emerge at high-symmetry points in crystals characterized by non-polar chiral point groups ($D_2$, $D_3$, $D_4$, $D_6$, $T$, $O$). Based on the symmetry constraints imposed by these point groups in a vector field, we study the general expression for the radial spin textures up to third order in momentum. Furthermore, we determine the high-symmetry points of the 45 non-polar chiral space groups supporting a radial spin texture. These two principles are used to screen materials databases for archetypes that go beyond the basic hedgehog radial spin texture. Among the selected materials we highlight the axion insulator candidate $\mathrm{Ta}_2 \mathrm{Se}_8\mathrm{I}$, the material proposed for dark matter detection $\mathrm{Ag}_3\mathrm{Au}\mathrm{Te}_2$ and heazlewoodite $\mathrm{Ni}_3\mathrm{S}_2$, a conventional metal predicted to exhibit current-induced spin polarization. We point out that the symmetry analysis proposed in this Letter is more general and extends to studying other vector properties in momentum space.
\end{abstract}


\maketitle


Many spin-related phenomena in condensed matter physics are understood in terms of the specific distribution in momentum space of $\textbf{S}_n(\textbf{k})=\langle \psi_n(\textbf{k}) \vert \hat{\textbf{S}} \vert \psi_n(\textbf{k})\rangle$, the spin expectation value of the Bloch wave function $\psi_n(\textbf{k})$ of band $n$ with momentum $\textbf{k}$. This quantity provides a spin texture (ST) in momentum space which is determined by the interplay between spin-orbit coupling (SOC) and crystal symmetries \cite{SAMOKHIN20092385, SAMOKHIN2019179}. The archetypal examples of such relationship are the effective $k \cdot p$ models of the Dresselhaus \cite{PhysRev.100.580} and Rashba \cite{Rashba_original, Bihlmayer_2015} spin-orbit couplings for non-centrosymmetric zinc blende and wurtzite crystal structures, respectively. However, since a $k \cdot p$ model depends on the crystal symmetries as well as the specific manifold of the Bloch states that are considered, it is a challenging task to obtain effective models to describe STs in a generic crystal. Such complex dependence on the details of the material is well illustrated by the existence of exotic Rashba-type spin textures with cubic momentum-dependent terms in the Bi/Ag(111) surface alloy~\cite{PhysRevB.85.075404} and rare-earth ternary materials~\cite{PhysRevLett.124.237202}. Recently, a new method based on topological quantum chemistry~\cite{Bradlyn2} has been proposed to address this problem. This method correlates the nature of the Bloch functions in a local basis around a multifold fermion to predict its ST~\cite{Bradlyn}. 

An alternative approach for obtaining information on the ST, or the momentum dependence of any vector field, consists in examining its transformation under the crystal symmetries. New STs have been identified on the basis of such symmetry analysis. For instance, the persistent ST, a uniform distribution of spins in momentum space, has been shown to be enforced by symmetry~\cite{Tao2018, PhysRevB.101.125207}. This type of ST also presents an exotic version with cubic momentum dependence around high-symmetry points with $C_{3h}$ and $D_{3h}$ point group (PG) symmetries \cite{PhysRevLett.125.216405}.  As highlighted in Ref.~\onlinecite{PhysRevB.104.104408}, the relevant symmetry operations that the vector field must satisfy change across the Brillouin zone (BZ) and are dictated by the PG symmetries of the wavevector around which the ST is evaluated.  

Radial spin distributions also emerge as new types of texture imposed by crystal symmetry \cite{PhysRevLett.114.206401}. Such STs appear around high-symmetry points in the BZ where spins are forced to point parallel to the momentum along several rotation symmetry axes. The lack of inversion and mirror symmetries, and the presence of three or more rotational axes are the necessary conditions for the radial STs. The only PGs that respect these conditions are the non-polar chiral $D_2$, $D_3$, $D_4$, $D_6$, $T$ and $O$.

Radial STs are commonly pictured either as a hedgehog configuration with the spins pointing parallel to the momentum along all the rotation symmetry axes or as a vector field where the spins point parallel and antiparallel to the momentum at different rotation symmetry axes \cite{chang2018topological, dil2020finding, Wei_Tan}. 
The first reported examples of materials hosting such STs are CoSi \cite{doi:10.7566/JPSCP.3.016019} and elemental Te \cite{PhysRevLett.114.206401}, two well-known chiral materials. Later, these STs were associated with the special Kramers-Weyl (KW) points located at time-reversal invariant momentum (TRIM) points in chiral materials. Experimental confirmation of radial STs via spin-resolved angle-resolved photoemission spectroscopy (ARPES) has been provided in Te at both TRIM \cite{PhysRevLett.125.216402} and non-TRIM high-symmetry points \cite{PhysRevLett.124.136404}, as well as in PtGa \cite{krieger2022parallel}. Theory predicted that materials with such spin configurations present a new magnetoelectric effect where a magnetization parallel to the current is induced \cite{yoda2015current, yoda2018orbital}. This effect has been experimentally confirmed in Te \cite{furukawa2017observation, PhysRevResearch.3.023111,calavalle2022gate}, ${\mathrm{CrNb}}_{3}{\mathrm{S}}_{6}$ \cite{PhysRevLett.124.166602} and chiral disilicides \cite{PhysRevLett.127.126602}. The advance opens a new route for the generation of long-range spin accumulation that can be used for spin manipulation by electrical means only \cite{He2021,calavalle2022gate,2203.05518}.

In this Letter, we propose an alternative route for classifying spin textures in momentum space. We systematically analyze the constraints imposed by the symmetries  with chiral non-polar little groups on a general vector field at high-symmetry points in momentum space. 
We provide a list of all the high-symmetry points (both TRIM and non-TRIM) from the 45 non-polar chiral space groups where it is possible to find a radial ST. Using examples of several known materials, we illustrate different types of radial STs according to the PG symmetry, the presence of time-reversal symmetry around the high-symmetry point, and the need to include higher-order momentum terms. 

The ST $\textbf{S}(\textbf{k}) = (S^x(\textbf{k}), S^y(\textbf{k}),S^z(\textbf{k}))$ around $\mathbf{k}_0$ (we omit the band index for simplicity), can be expressed as the Taylor expansion of its components
\begin{equation}
\label{eq:1}
S^{\alpha}(\textbf{q}) = \sum_{i,j,k} S^{\alpha}_{ijk} q_x^i q_y^j q_z^k , \,\,\,\,\, \alpha = x, y, z,
\end{equation}
where $\textbf{q}=\textbf{k}-\textbf{k}_0$ is the momentum measured from the expansion point $\textbf{k}_0$, typically a high-symmetry point. The real coefficients $S^{\alpha}_{ijk}$ are not arbitrary and must satisfy the symmetry requirements of the vector field around $\textbf{k}_0$, \emph{i.e.,} the symmetries of the group of $\mathcal{G}_{\textbf{k}_0}$. Since translations do not affect momentum space properties, only PG symmetries determine how a vector or pseudo-vector transforms. Let $g$ be the matrix representation of a symmetry of $\mathcal{G}_{\textbf{k}_0}$. A pseudo vector must transform as
\begin{equation}
\textbf{S}(\textbf{q})  = \mathrm{det}(g)\, g\textbf{S}(g^{-1}\textbf{q}), \,\,\,\,\, g \in \mathcal{G}_{\mathbf{k}_0}.
\label{eq:2}
\end{equation}
In the case of non-polar chiral PGs only proper rotations with $\mathrm{det}(g)=1$ are present, then pseudo-vectors transform as  normal vectors. By imposing the symmetry constraints of each non-polar chiral PG, we determine which $S^{\alpha}_{ijk}$ coefficients vanish up to the third order. Table~\ref{tab:A} summarizes the most general expressions of the vector field compatible with each non-polar chiral PG. These expressions describe the ST up to a global function $f(q)$ that depends on the modulus of $\textbf{q}$. Let $\textbf{S}(\textbf{q})$ be a vector from Table \ref{tab:A} that satisfies Eq.~(\ref{eq:2}), then $\textbf{S'}(\textbf{q})=f(q)\textbf{S}(\textbf{q})$ satisfies Eq.~(\ref{eq:2}) as well. Real coefficients $A_i$, $B_i$ and $C_i$ correspond to non-zero parameters of the linear, quadratic and cubic terms, respectively. .

\begin{table}[htb]
\caption{\label{tab:A} Expansion terms of the spin texture up to the third order for the non-polar chiral point groups.}
\resizebox{0.485\textwidth}{!}{
\begin{tabular}{ c c >{\columncolor{Gray}}c c >{\columncolor{Gray}}c  } 
 \hline 
 \hline
 \xrowht{7pt} Point group & $\alpha$ & $1^{st}$ & $2^{nd}$ & $3^{rd}$ \\  
 \hline
 \hline
\xrowht{7pt} \multirow{3}{*}{$D_2$}& $x:$ & $A_1 q_{x}$ & $B_1 q_{y}q_{z}$ &  $C_1 q_{x}^3 + C_2 q_{x}q_{y}^2 + C_3  q_{x}q_{z}^2$ \\ \hhline{~|-|-|-|-|} 
\xrowht{7pt}                       & $y:$ &  $A_2 q_{y}$ & $B_2 q_{x}q_{z}$ &  $C_4 q_{y}^3 + C_5 q_{y}q_{z}^2 + C_6  q_{y}q_{z}^2$ \\  \hhline{~|-|-|-|-|} 
\xrowht{7pt}                      & $z:$&  $A_3 q_{z}$ & $B_3 q_{x}q_{y}$ &  $C_7 q_{z}^3 + C_8 q_{z}q_{x}^2 + C_9  q_{z}q_{y}^2$\\         
\hline                                                                               
\hline             
\xrowht{7pt} \multirow{3}{*}{$D_4$}& & $A_1 q_{x}$ & $B_1 q_{y}q_{z}$&  $C_1 q_{x}^3 + C_2 q_{x}q_{y}^2 + C_3 q_{x}q_{z}^2$ \\ \hhline{~|-|-|-|-|} 
\xrowht{7pt}                       & &  $A_1 q_{y}$ & -$B_1 q_{x}q_{z}$&  $C_1 q_{y}^3 + C_2 q_{y}q_{x}^2 + C_3 q_{y}q_{z}^2$ \\ \hhline{~|-|-|-|-|} 
\xrowht{7pt}                       &  &$A_2 q_{z}$ &  & $C_4 q_{z}^3 + C_5 q_{z} (q_{x}^2+ q_{y}^2)$\\         
\hline                                                                     
\hline                                                                       
\xrowht{7pt} \multirow{3}{*}{$D_3$} &  &$A_1 q_{x}$ & -$2 B_1 q_{x}q_{y} + B_2 q_{y}q_{z} $  &  $C_1 (q_{x}^3 + q_{x}q_{y}^2) + C_2 q_{x}q_{z}^2 $ \\ \hhline{~|-|-|-|-|} 
\xrowht{7pt}                        &  &$A_1 q_{y}$ & $B_1(q_{y}^2 - q_{x}^2)-B_2 q_{x}q_{z}$ & $C_1 (q_{y}^3 + q_{y}q_{x}^2) + C_2 q_{y}q_{z}^2 $  \\ \hhline{~|-|-|-|-|} 
\xrowht{7pt}                        &  &$A_2 q_{z}$ &  &  $C_3(q_x^3-3q_xq_y^2) + C_4 q_{z} (q_{x}^2+q_{z}^2) + C_5 q_{z}^3$\\   
\hline                                                               
\hline
\xrowht{7pt} \multirow{3}{*}{$D_6$} &  &$A_1 q_{x}$ & $ B_1 q_{y}q_{z} $  &  $C_1 (q_{x}^3 + q_{x}q_{y}^2) + C_2 q_{x}q_{z}^2 $ \\ \hhline{~|-|-|-|-|}  
\xrowht{7pt}                        &  &$A_1 q_{y}$ & -$ B_1 q_{x}q_{z}$ &  $C_1 (q_{y}^3 + q_{y}q_{x}^2) + C_2 q_{y}q_{z}^2 $ \\ \hhline{~|-|-|-|-|} 
\xrowht{7pt}                        &  &$A_2 q_{z}$ &  &  $C_3 q_{z} (q_{x}^2+q_{z}^2) + C_4 q_{z}^3$ \\       
\hline                                                      \hline
\xrowht{7pt} \multirow{3}{*}{$T$} & &$A_1 q_{x}$ & $ B_1 q_{y}q_{z} $  & $C_1 q_{x}^3 + C_2 q_x q_{y}^2 +  C_3 q_x q_{z}^2$ \\ \hhline{~|-|-|-|-|}  
\xrowht{7pt}                      & &$A_1 q_{y}$ & $ B_1 q_{x}q_{z}$ & $C_1 q_{y}^3 + C_2 q_y q_{z}^2 +  C_3 q_y q_{x}^2$ \\ \hhline{~|-|-|-|-|} 
\xrowht{7pt}                      & &$A_1 q_{z}$ & $ B_1 q_{x}q_{y}$ & $C_1 q_{z}^3 + C_2 q_z q_{x}^2 +  C_3 q_z q_{y}^2$\\     
\hline                                                          
\hline
\xrowht{7pt} \multirow{3}{*}{$O$} & & $A_1 q_{x}$ &  &  $C_1 q_{x}^3 + C_2 q_x (q_{y}^2 + q_{z}^2)$ \\ \hhline{~|-|-|-|-|}  
\xrowht{7pt}                      &  &$A_1 q_{y}$ &  &  $C_1 q_{y}^3 + C_2 q_y (q_{x}^2 + q_{z}^2)$ \\  \hhline{~|-|-|-|-|} 
\xrowht{7pt}                      &  &$A_1 q_{z}$ &  &  $C_1 q_{z}^3 + C_2 q_z (q_{x}^2 + q_{y}^2)$ \\    
\hline
\hline
\end{tabular}
}
\end{table}

\begin{figure}[b]
\includegraphics[]{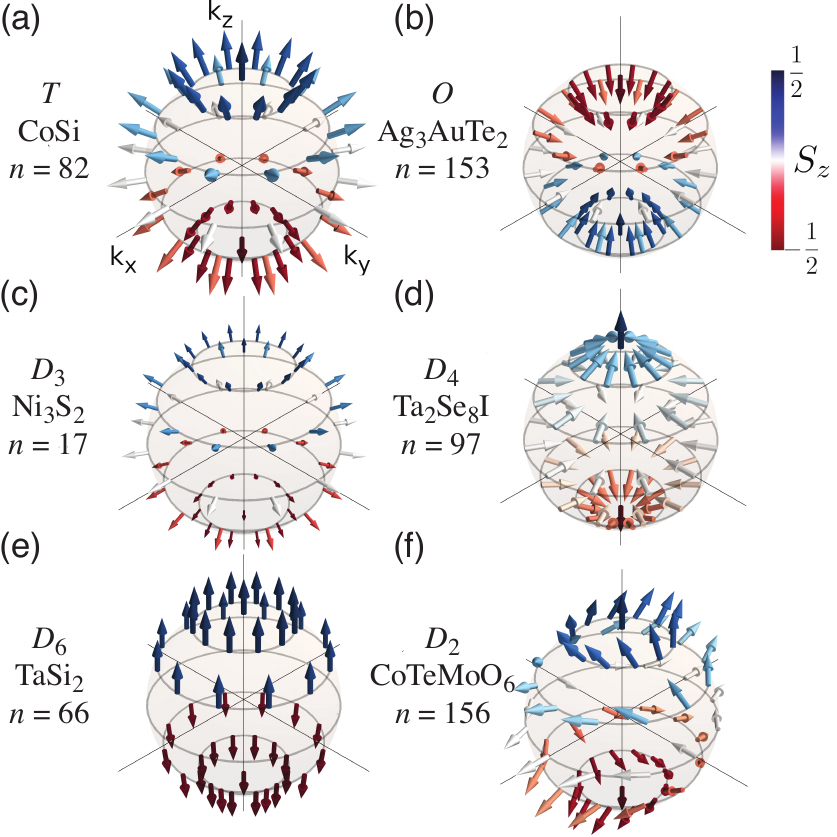}
\caption{Examples of radial spin textures with linear momentum dependence. Respective materials and band indices are labeled.
In all the cases the sphere surrounds the $\Gamma$ point.
\label{Fig1}}
\end{figure}

In order to identify all possible points in momentum space that can support radial STs in generic crystals, we determine the high-symmetry points, for the 45 eligible space groups, with non-polar chiral PGs. Following the double-value representation tables in Ref.~\onlinecite{bradley2010mathematical}, we identify 149 high-symmetry points with a non-polar chiral PG. These points are listed in Table~I 
of the Supplemental Material \cite{SM}. Only the points not intersected by high-symmetry lines or planes with symmetry-enforced degeneracies are selected, otherwise the continuity criteria required for performing Taylor expansion is not satisfied. 

For a given space group with PG $\mathcal{G}$, the high-symmetry points of the corresponding BZ have a point group $\mathcal{G}_{\textbf{k}_0} \subseteq \mathcal{G}$. When $\mathcal{G}_{\textbf{k}_0}$ is a subset of $\mathcal{G}$, the star of the high-symmetry point $\textbf{k}_0$ contains more than one element. At each of these non-equivalent points the spin vector field bears a different Taylor expansion, but the coefficients $S^{\alpha}_{ijk}$ of the vector field expansion at non-equivalent points of the star are related by the complementary symmetry operations that form subgroup $\mathcal{Q}_\textbf{k}=\mathcal{G}-\mathcal{G}_{\textbf{k}_0}$. These additional relations among $S^{\alpha}_{ijk}$ at different high-symmetry points of the star can introduce new constraints on the vector field and even reduce the number of constants required to describe such field. The details of constraints imposed by the PG of the space group are given in the Supplemental Material \cite{SM}.

\begin{figure}
\includegraphics[]{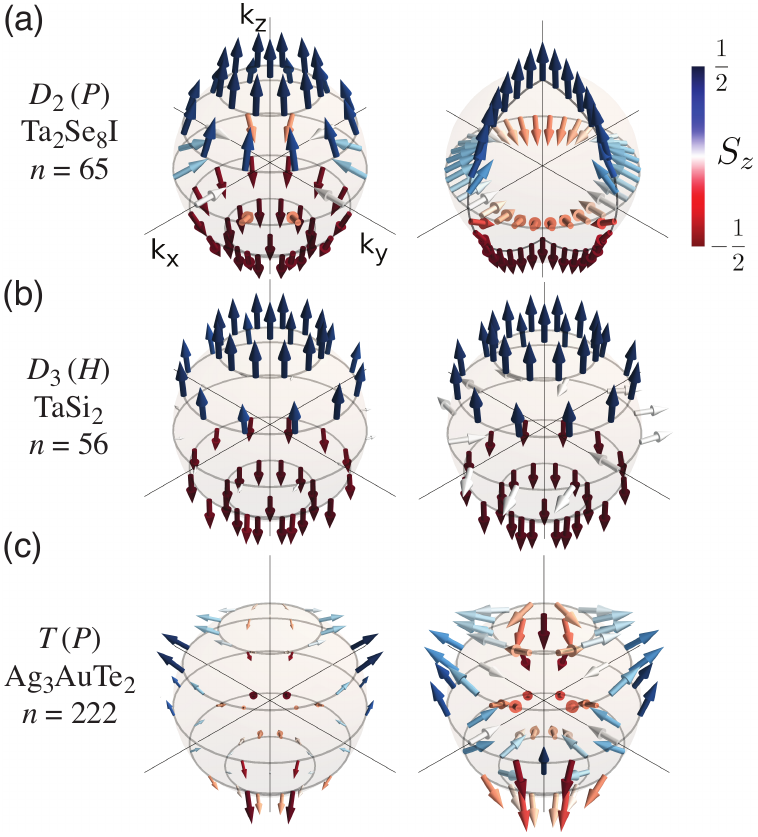}%
\caption{Examples of radial spin textures with quadratic momentum dependence. The corresponding high-symmetry points in the Brillouin zone are indicated in parentheses.   
Left panels are the computed spin textures and right panels are the normalized spin texture for the same band.
\label{Fig2}}
\end{figure}

Next, we illustrate the diversity of spin textures around  high-symmetry points for each non-polar chiral PG \footnote{We compute, using density functional theory, spin textures around small spheres of radius 0.005 $2\pi/a$, where $a$ is the lattice constant of the material, surrounding the high-symmetry points. See the Supplemental Material~\cite{SM} for details of calculations.} using a selection of chiral materials listed in Table~\ref{tab:B}. These materials are chosen according to the  following criteria. Both $\mathrm{CoSi}$ and $\mathrm{TaSi}_2$ highlight the complexity of the radial ST in well-known chiral crystals that were assumed to host a simple hedgehog spin configuration. We have also included the axion insulator candidate $\mathrm{Ta}_2 \mathrm{Se}_8\mathrm{I}$ \cite{Gooth2019} and the material proposed for dark matter detection $\mathrm{Ag}_3\mathrm{AuTe}_2$ \cite{dark} to emphasize the possibility of simultaneous manifestation of several exciting electronic phenomena in these novel materials. Finally, in order to complete the list of STs generated by the different non-polar chiral PGs, we screened the TopoMat database \cite{TopoMat} with the help of Table~I
in the Supplemental Material \cite{SM} to find new materials that can host radial STs. We selected heazlewoodite $\mathrm{Ni}_3\mathrm{S}_2$ \cite{PhysRevB.54.13542} and the non-magnetic phase of $\mathrm{CoTeMoO}_6$ as examples of the $D_3$ and $D_2$ PG symmetries. Special interest must be paid to heazlewoodite, a conventional metal that is susceptible to exhibit the chirality-induced spin polarization. In all cases, we choose STs produced by the bands that are the most representatives of all scenarios, bearing in mind that these are not isolated cases. 

\begin{table}[b]
\caption{\label{tab:B} Selected materials and their space groups illustrating the spin texture imposed by each non-polar point group.}
\begin{tabular}{ c c c } 
 \hline\hline
\xrowht{7pt} Point group & TRIM & Non-TRIM \\ 
 \hline
\xrowht{7pt} {$D_2$} & $\mathrm{CoTeMoO}_6$ (SG 18)  &  {$\mathrm{Ta}_2 \mathrm{Se}_8\mathrm{I}$ (SG 97)} \\
\hline
\xrowht{7pt} $D_3$  & $\mathrm{Ni}_3 \mathrm{S}_2$ (SG 155) & $\mathrm{TaSi}_2$ (SG 180)  \\
\hline
\xrowht{7pt} $D_4$ & $\mathrm{Ta}_2 \mathrm{Se}_8\mathrm{I}$ (SG 97)  & - \\
\hline
\xrowht{7pt} $D_6$ &  $\mathrm{TaSi}_2$ (SG 180)  & -\\
\hline
\xrowht{7pt} $T$ & CoSi (SG 198) & $\mathrm{Ag_3AuTe}_2$ (SG 214) \\
\hline
\xrowht{7pt} $O$ & $\mathrm{Ag}_3\mathrm{AuTe}_2$ (SG 214) & - \\
\hline
\hline
\end{tabular}
\end{table}

So far, the spin textures of non-polar chiral PGs have been studied only up to linear order. This simplification gave rise to the notion of a hedgehog vector field. However, hedgehog or purely radial or hedgehog STs are only present around high-symmetry points with $T$ and $O$ PGs, while for the rest of PGs the textures cannot be purely radial. Focusing on the linear terms in Table~\ref{tab:A}, purely radial STs of the $T$ and $O$ PGs are described by a vector field with all three components having the same coefficient $A_1$. Examples of such purely radial STs are illustrated in Figs.~\ref{Fig1}(a) and \ref{Fig1}(b) for CoSi ($T$) and $\mathrm{Ag_3AuTe}_2$ ($O$) at the $\Gamma$ point.

As the number of rotation axes of the PGs is reduced, the number of parameters describing the ST increases, and non-radial components along the directions other than the rotation axes emerge. For the $D_3$, $D_4$ and $D_6$ PGs, two different parameters $A_1 \ne A_2$ are required. For the $D_2$ PG each spin component has a different coefficient $A_1$, $A_2$ and $A_3$. Examples of possible STs that 
correspond to specific bands at the $\Gamma$ point in $\mathrm{Ni}_3\mathrm{S}_2$, $\mathrm{Ta}_2\mathrm{Se}_8\mathrm{I}$, $\mathrm{TaSi}_2$ and $\mathrm{CoTeMoO}_6$ are shown in Figs.~\ref{Fig1}(c)--\ref{Fig1}(f), respectively. 
While Fig.~\ref{Fig1}(c) shows an almost purely radial ST due to $A_2$ being slightly larger than $A_1$, the rest of examples [Figs.~\ref{Fig1}(d)--\ref{Fig1}(f)] deviate strongly from the ideal hedgehog configurations. This is due to $|A_1|\approx|A_2|$ but $\mathrm{sign}(A_1) \ne \mathrm{sign}(A_2)$,  $A_2 \gg A_1$ and $A_1 \ne A_2 \ne A_3$, respectively.  We would like to emphasize the case shown in Fig.~\ref{Fig1}(e), in which $A_2$ is significantly larger than $A_1$ and a near-persistent ST is created between the north and south hemisphere. 

\begin{figure*}
\includegraphics[]{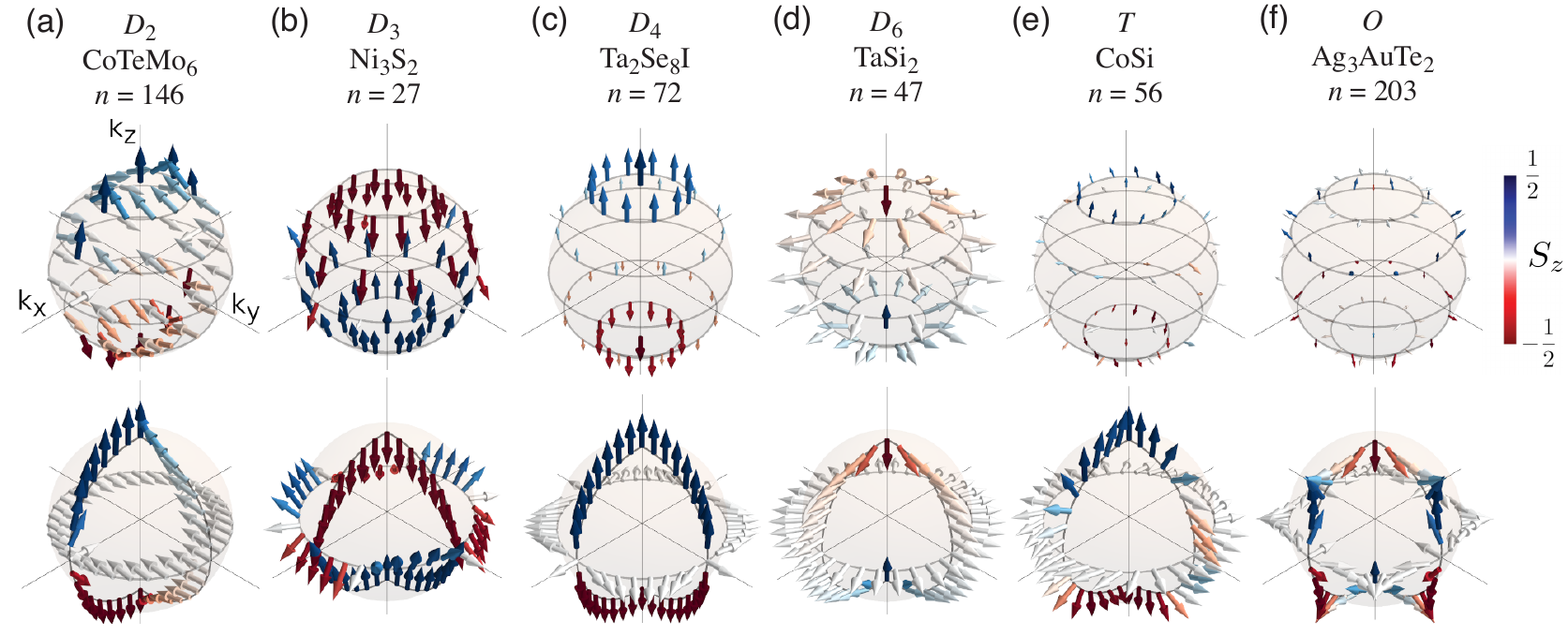}%
\caption{Examples of radial spin textures with cubic momentum dependence.
In all the cases the sphere surrounds the $\Gamma$ point. Top panels are the computed spin texture and bottom panels the normalized spin texture for the same band.
\label{Fig3}}
\end{figure*}

Although radial STs have been associated with KW points \cite{chang2018topological} and multifold band degeneracies at TRIM points, such configurations also appear at non-TRIM points. The seminal work on elemental Te was devoted to the study of the band structure and the ST around the $H$ point, a non-TRIM point at which the two highest valence bands are non-degenerate. There is, however, a fundamental difference since at TRIM points the quadratic terms of Eq.~(\ref{eq:1}) are forced to be zero due to time-reversal symmetry, while this condition is not present at non-TRIM points. Non-TRIM high-symmetry points are less common, and we identified a total of 24 distributed in the $D_2$, $D_3$ and $T$ point groups. In order to illustrate the appearance of radial STs due to leading quadratic momentum terms at non-TRIM points, we selected three materials from the TRIM column in Table~\ref{tab:B} that also have non-TRIM points in their BZ. These are $\mathrm{Ta}_2\mathrm{Se}_8\mathrm{I}$, $\mathrm{Ta}\mathrm{Si}_2$ and $\mathrm{Ag}_3\mathrm{Au}\mathrm{Te}_2$ for the $D_2$, $D_3$ and $T$ point groups, respectively, and the corresponding STs are shown in Figs.~\ref{Fig2}(a)--\ref{Fig2}(c). 
In order to highlight the details of the STs, we divided each panel into two parts: left part shows the computed ST, while the right displays the normalized ST ($\textbf{S}_n(\textbf{k})/|\textbf{S}_n(\textbf{k})|$) for the same band. In the following, we describe the contribution of the different quadratic terms to the three cases.  
Figure~\ref{Fig2}(a) shows the ST of $\mathrm{Ta}_2\mathrm{Se}_8\mathrm{I}$ for a band around the $P$ point. The quadratic behaviour of the spin is evident at the equator $q_z=0$, where the $B_3 q_x q_y$ term of the $S_z$ component is manifested. Similarly, the selected ST of $\mathrm{Ta}\mathrm{Si}_2$ in Fig.~\ref{Fig2}(b) is governed by the $B_1 q_x q_y$ and $B_1(q_y^2-q_x^2)$ terms of the $S_x$ and $S_y$ components, respectively. Finally, the case of $\mathrm{Ag}_3\mathrm{Au}\mathrm{Te}_2$ at the $P$ point [Fig.~\ref{Fig2}(c)] shows a competition between the linear and quadratic terms with $\mathrm{sign}(A_1) \neq \mathrm{sign}(B_1)$. 

Furthermore, Fig.~\ref{Fig2}(a) illustrates an example of ST at a high-symmetry point $\textbf{k}_0$ with a PG $\mathcal{G}_{\textbf{k}_0}$, a subgroup of space group $\mathcal{G}$, being conditioned by the subgroup $\mathcal{Q}_\textbf{k}=\mathcal{G}-\mathcal{G}_\textbf{k}$.  According to Table~\ref{tab:A}, four parameters determine the ST up to quadratic order at the equator encircling a point with the $D_2$ point group. As a consequence, low-symmetry STs are possible, but in $\mathrm{Ta}_2\mathrm{Se}_8\mathrm{I}$ the $C_4$ rotation symmetry of the crystal point group $\mathcal{G}$ relates the vector field at non-equivalent points $P$. As a result, at these points $A_1 = A_2$ and $B_1 = B_2$ resulting in a more symmetric ST.

In general, the complexity of ST around high-symmetry points with non-polar chiral PGs goes beyond the linear and quadratic terms. For each material in Table~\ref{tab:B} it is possible to find several bands, for which STs require cubic terms to describe momentum dependence. Figure~\ref{Fig3} shows some examples that we found around the $\Gamma$ point of each material. The top panels show the computed ST, while the bottom panels illustrate its normalized vectors. In the example of the $D_2$ point group symmetry [Fig.~\ref{Fig3}(a)], the $S_z$ component is dominated by the $q_z q_x^2$ term since the spin only points almost parallel to $k_z$ as we move along the $k_x$-$k_z$ plane. Figure \ref{Fig3}(b) illustrates an example for the $D_3$ point group with $S_z$ being governed by the $C_3(q_x^3-3q_xq_y^2)$ term close to the equator of the sphere. The ST shown in Fig.~\ref{Fig3}(c) corresponds to the $D_4$ point group. In this case, the spin texture near the equator is the result of a competition between the cubic terms $C_1$ and  $C_2$ with $\mathrm{sign}(C_1) = \mathrm{sign}(C_2)$ but $C_2 > C_1$. In Figs.~\ref{Fig3}(d) and \ref{Fig3}(e), the nature of the cubic spin texture is revealed by a fast change in the orientation of ST on the sphere. In the case of the $D_6$ point group [Fig.~\ref{Fig3}(d)], the $S_z$ component is governed by the $q_z^3$ term because it rapidly decays to zero with $k_z$. For the $T$ point group symmetry [Fig.~\ref{Fig3}(e)], the $q_x=0$ and $q_y=0$ planes are asymmetric, therefore $\mathrm{sign}(C_2) \ne \mathrm{sign}(C_3)$. Finally, the selected example for the $O$ point group [Fig.~\ref{Fig3}(f)] presents a ST with a high winding for each high-symmetry plane. Such ST is generated by the competition of the  $A_1$ and $C_1$ term with $\mathrm{sign}(A_1)\ne \mathrm{sign}(C_1)$.

To summarize, we have studied the radial STs generated by the non-polar chiral PGs. Based on the symmetry transformation of a pseudo-vector under the operations of these PGs, we provide the most general expression up to the third order in momentum expansion. We show that beyond the simplicity that the name radial spin texture alludes to, many exotic STs with a non-radial behaviour appears. 
We also provide a comprehensive list of all the high-symmetry points where such STs can be found among the 45 non-polar chiral space groups. These results give a simple description of the radial STs based on the PG symmetries of the high-symmetry points that can be used for predict the radial ST and correlate it with the current-induced spin polarization. Our results allow for straightforward screening of materials databases in order to find novel compounds with radial STs.
Finally, the symmetry analysis provided in this Letter is not only valid for the spin texture, but for all the other vector properties in momentum space. 

\begin{acknowledgments}
D.G.-M. and O.V.Y. acknowledge support by the NCCR MARVEL, a National Centre of Competence in Research, funded by the Swiss National Science Foundation (grant number 182892). First-principles calculations were performed at the Swiss National Supercomputing Centre (CSCS) under project No.~s1146. D.G.-M. thanks the María Zambrano Programme at the University of Alicante founded by the European union-Next Generation EU as well as J.~J.~Palacios for insightful discussions. This study forms part of the Advanced Materials program and was supported by the Spanish MCIN with funding from European Union NextGenerationEU and by Generalitat Valenciana through MFA/2022/045.
\end{acknowledgments}


%

\end{document}